\documentclass[12pt]{article}
\usepackage{graphicx}
\def\ltap{\raisebox{-.4ex}{\rlap{$\sim$}} \raisebox{.4ex}{$<$}}
\def\gtap{\raisebox{-.4ex}{\rlap{$\sim$}} \raisebox{.4ex}{$>$}}

\def\sinthlsp {${\rm sin}^2 \theta_W^{\ell\ {\rm eff}}\ $}

\def\tpsp {$T^{\prime}\ $}
\def\mtp {$m_{t^{\prime}}$}
\def\mtpsp {$m_{t^{\prime}}\ $}
\def\sth34{$|{\rm sin}\theta_{34}|$}
\def\sth34sp{$|{\rm sin}\theta_{34}|\ $}

\def\bpsp {$b^{\prime}\ $}

\def\tpsp {$t^{\prime}\ $}

\def\alr {$A_{LR}$}
\def\alrsp {$A_{LR}\ $}
\def\afbb {$A_{FB}^b$}
\def\afbbsp {$A_{FB}^b\ $}

\def\afbc {$A_{FB}^c$}

\def\qfb {$Q_{FB}$}

\def\mh {$m_H$}
\def\mhsp {$m_H$\ }
\def\chisqsp {$\chi^2$\ }

\def\th34{$\theta_{34}$}
\def\th34sp{$\theta_{34}$\ }
\def\journal{\topmargin 0.0in   \oddsidemargin 0in
        \headheight 0pt \headsep 0pt
        \textwidth 6.5in 
\textheight 9in 
        \marginparwidth 1.5in
        \parindent 2em
        \parskip .5ex plus .1ex         \jot = 1.5ex}
\journal

\begin{document}
\begin{titlepage}

\noindent April 22, 2009 \hfill arXiv:0904.3570 [hep-ph]\\
\noindent Rev. April 27, 2009\\

\begin{center}

\vskip .5in

{\large Bounding CKM Mixing with a Fourth Family}

\vskip .5in

Michael S. Chanowitz

\vskip .2in

{\em Theoretical Physics Group\\
     Lawrence Berkeley National Laboratory\\
     University of California\\
     Berkeley, California 94720}
\end{center}

\vskip .25in

\begin{abstract}

CKM mixing between third family quarks and a possible fourth family is
constrained by global fits to the precision electroweak data.  The
dominant constraint is from nondecoupling oblique corrections rather
than the vertex correction to $Z \to \overline bb$ used in previous
analyses. The possibility of large mixing suggested by some recent
analyses of FCNC processes is excluded, but 3-4 mixing of the same
order as the Cabbibo mixing of the first two families is allowed.

\end{abstract}

\end{titlepage}

\newpage

\renewcommand{\thepage}{\arabic{page}}
\setcounter{page}{1}

{\it \noindent \underline{Introduction}}

A fourth family of quarks and leptons is an obvious extension of the
Standard Model (SM) that will be investigated at the LHC.\cite{hs} If
a fourth family is discovered, it is likely to have consequences at
least as profound as those that have emerged from the discovery of the
third family. The necessarily heavy neutrino mass, $m_{\nu_4}>m_Z/2$,
would be surprising, but without a theory of neutrino masses we are
not really in a position to judge; if a fourth family were discovered,
it would instantaneously refocus the effort to understand neutrino
masses (see for instance \cite{hill,bhnu}).  A fourth family is consistent
with precision electroweak (EW) data,\cite{he,okun,kribs} and can
remove\cite{okun} the tension between the SM fit and the LEP II lower
bound on \mhsp that arises if the $3.2\sigma$ discrepancy between
hadronic and leptonic determinations of \sinthlsp turns out to be the
result of underestimated systematic error.\cite{mcmh} The SM augmented
with a fourth family is consistent with $SU(5)$ gauge coupling
unification without supersymmetry.\cite{pqh} Electroweak baryogenesis
might be viable with four families,\cite{ewbg} although it is not in
the SM with only three. Since the plausible parameter space includes
the strong coupling region determined by perturbative
unitarity,\cite{cfh} $m_{Q_4}\, \gtap\, 550$ GeV, a heavy fourth
family could naturally play a role in the dynamical breaking of
electroweak symmetry.\cite{hill,3ewsb} Even if fourth family quarks are
very heavy, e.g., $m_Q\, \gtap\, 1$ TeV, and difficult or 
impossible to observe directly, they will give rise to a large
nonresonant signal for production of longitudinally polarized $Z$
boson pairs from $gg \to ZZ$, that could be seen at the LHC with
$5\sigma$ significance over backgrounds with only O(10) fb$^{-1}$ of
integrated luminosity.\cite{mczz}

Motivated initially by an interesting study of the FCNC constraints on
a unitary $4 \times 4$ CKM matrix by Bobrowski {\it et
  al.},\cite{blrr} we have studied the constraint on 3-4 family CKM
mixing that can be obtained from precision electroweak data. They
found, in addition to the expected small angle solutions, that
surprisingly large mixing between the third and fourth family quarks
is also allowed. They exhibit fits with $|V_{tb}|/|V_{tb}^{SM3}|$ as
small as 0.73, corresponding to $|V_{t^{\prime}b}|$ as large as
$\simeq 0.63$, just at the edge of the 95\% allowed region for
$|V_{tb}|$ determined from single top production.\cite{pdg} We find
however that these fits are decisively excluded by the precision EW
data and present the EW constraints on 3-4 family CKM mixing for a
range of fourth family masses favored by the EW data.  Although the
large-mixing FCNC fits are excluded, the EW constraints do allow 3-4
CKM mixing of the same order as the Cabbibo mixing of the first two
families. Our results are also inconsistent with large 3-4 mixing
parameters obtained in another recent study\cite{herrera} and
constrain proposals to explain the CP anomalies suggested by B meson
data.\cite{hou,soni}

In the presence of 3-4 CKM mixing there are two nondecoupling
radiative corrections to the precision EW observables with quadratic
sensitivity to heavy fourth-family fermion masses: the $\rho$
parameter correction,\cite{tini,cfh} (AKA the oblique parameter
$\alpha T$\cite{pt}) and the $Zb\overline b$ vertex
correction.\cite{zbb} In the SM both are proportional to $G_Fm_t^2$ at
one loop order. In the four-family model they give rise to corrections
proportional to $|V_{t^{\prime}b}|^2 m_{t^{\prime}}^2$ and, in the
case of the oblique corrections, there are also large corrections
proportional to $|V_{tb^{\prime}}|^2 m_{b^{\prime}}^2$ if
$m_{b^{\prime}}^2\gg m_t^2$.

Previous consideration of the precision EW constraint on 3-4 CKM
mixing focused on the effect of the vertex correction on $R_b=
\Gamma(Z\to \overline bb)/\Gamma(Z\to {\rm
  hadrons})$,\cite{yanir,alwall} which was used as a constraint in
subsequent FCNC studies (e.g., \cite{hou,soni}). However the oblique
corrections are of precisely the same order and actually provide the
most important constraint.  To obtain a valid bound it is essential to
reevaluate the global EW fit in the new physics model, since the not
infrequently followed practice of using just the magnitude of shifts
from the values in the SM fit does not take into account the
possibility that a global fit incorporating the new physics may have
its \chisqsp minimum at significantly different values of the SM
parameters, e.g., $m_t$ and, especially, \mh.\footnote{Alwall {\it et
    al.} consider the oblique constraints and remark that they are
  sensitive to \mh. They do not perform a global fit, which would
  incorporate the \mhsp dependence, and instead rely on $R_b$ for
  their strongest precision EW constraint.} This in fact occurs in the
results presented below. In the fits that establish the 95\% CL limits
(e.g., tables 2 and 5) $R_b$ is only $1\sigma$ from its experimental
value and does not contribute to the constraint.\footnote{The shift in
  $R_b$ is dominated by the nondecoupling vertex correction. The
  oblique corrections to $\Gamma(Z\to {\rm hadrons})$ and $\Gamma(Z\to
  \overline bb)$ are significantly larger but cancel in the
  ratio. $\Gamma(Z\to {\rm hadrons})$ and $\Gamma(Z\to \overline bb)$
  are not included in the fits but can be derived from combinations of
  observables that are.}  In addition, global fits are less
susceptible to statistical fluctuation or systematic uncertainty than
a constraint based on a single observable. For instance, the $2\sigma$
upper limit on 3-4 mixing obtained from the (negative) shift in $R_b$
from its SM value would have been significantly weaker if the SM value
for $R_b$ were $0.7\sigma$ above the experimental value rather than
$0.7\sigma$ below.

In the following sections we briefly review the nondecoupling
corrections and present the bounds on the mixing angle that follow
from the global fits.  Because the heaviest quark masses considered
are at the threshold of the strong coupling region, we use the two
loop correction to the $\rho$ parameter as a guide to the
applicability of perturbation theory and the accuracy of the results.

{\it \noindent \underline{Nondecoupling corrections}}

For vanishing CKM angles, the one loop correction to the $\rho$
parameter from a heavy fermion doublet $(f_1,f_2)$ is\cite{tini,cfh}
\begin{eqnarray}
\delta \rho= N_C\frac{\alpha}{8\pi x_W(1-x_W)}F_{12}
\end{eqnarray}
where $N_C=1,3$ for leptons and quarks respectively, $x_W= {\rm
  sin}^2\theta_W$, and $F_{12}$ is 
\begin{eqnarray}
   F_{12}= \frac{x_1 +x_2}{2}
              -\frac{x_1x_2}{x_1-x_2}{\rm ln}\frac{x_1}{x_2}
\end{eqnarray}
with $x_i = m_i^2/m_Z^2$. To study the EW constraints on the 
large-mixing parameter sets of \cite{blrr} it suffices to assume a
block-diagonal form for the 3-4 CKM submatrix,\footnote{We have
  verified explictly for the large-mixing parameter sets of
  \cite{blrr} that this approximation is valid to better than 1\% for
  the diagonal matrix elements and to better than 3\% for the
  off-diagonal ones. } which is then characterized by a single angle,
$\theta_{34}$, with $|V_{tb}|=|V_{t^{\prime}b^{\prime}}|= c_{34}$ and
$|V_{t^{\prime}b}|= |V_{tb^{\prime}}|= s_{34}$, where
$c_{34}\equiv{\rm cos}\theta_{34}$ and $s_{34}\equiv{\rm
  sin}\theta_{34}$. The oblique correction $T$ from the fourth family
is then
\begin{eqnarray}
T_4= \frac{1}{8\pi x_W(1-x_W)}\left\{ 3\left[
     F_{t^{\prime}b^{\prime}} +s_{34}^2(F_{t^{\prime}b} 
     +F_{tb^{\prime}} -F_{tb} -F_{t^{\prime}b^{\prime}})\right] 
     + F_{l_4\nu_4}\right\}.
\end{eqnarray}
The term $-s_{34}^2F_{tb}$ is the decrease from the three-family SM
$t\overline b$ contribution to the $W$ boson vacuum
polarization. $F_{t^{\prime}b}$ is the largest term and puts the
strongest constraint on $\theta_{34}$. Equation (3) is easily obtained
following the derivation for $V_{tb}=1$ in the second paper cited in
\cite{cfh}; for $V_{tb}\neq 1$ the GIM mechanism and the custodial
$SU(2)$ together ensure that the divergences cancel between the $W$
and $Z$ vacuum polarization terms, leaving the finite correction in
(3).

We also need the oblique correction $S$ for the fourth family quark
doublet, which to a very good approximation\cite{he} is given by
\begin{eqnarray}
S_4= \frac{N_C}{6\pi}\left(1 -\frac{1}{3}{\rm ln}\frac{x_1}{x_2}\right).
\end{eqnarray}
For the lepton masses considered below, the leptonic contribution to
$S$ is negligible.

Finally there is the nondecoupling correction to the
$Zb\overline b$ interaction of the left handed $b$ quark, which arises in
the SM from one loop vertex corrections containing the $t$ quark and
the $W$ boson.  In our notation the interaction Lagrangian is
\begin{eqnarray}
{\cal L}=\frac{g}{{\rm cos}\theta_W}g_{bL}\overline b_L {\not\! Z} b_L
\end{eqnarray}
where in the SM $g_{bL} = -\frac{1}{2} +\frac{1}{3}x_W$. In the SM the
one loop, nondecoupling vertex correction from the $t+W$ loop graphs
is\cite{zbb}
\begin{eqnarray}
\delta^V g_{bL}^{\rm SM} = \frac{\alpha}{16\pi x_W(1-x_W)}\, 
                                \frac{m_t^2}{m_Z^2}
\end{eqnarray}
The one loop correction from 3-4 CKM mixing is then
\begin{eqnarray}
\delta^V g_{bL}^{3-4} = s_{34}^2\, \frac{\alpha}{16\pi x_W(1-x_W)}\, 
     \left(\frac{m_{t^{\prime}}^2}{m_Z^2} - \frac{m_t^2}{m_Z^2}\right)
\end{eqnarray}
where again the last term accounts for the decrease of the SM top 
quark correction.

{\it \noindent \underline{Bounds on $\theta_{34}$}}

Like Bobrowski {\it et al.}\cite{blrr} we focus on parameters for the
fourth family shown by Kribs {\it et al.}\cite{kribs} to be favored by
the precision EW data. In addition to the three parameter sets
identified by Bobrowski {\it et al.}, shown in table 1, we survey
other \tpsp masses between 300 and 600 GeV.  Following \cite{okun} and
\cite{kribs} we confirm for $s_{34}=0$ that \chisqsp is minimized for
$|m_{t^{\prime}} - m_{b^{\prime}}| \sim 45 - 75\, {\rm GeV}$ and set
the \bpsp mass to
\begin{eqnarray}
m_{b^{\prime}}=m_{t^{\prime}} -55\, {\rm GeV}.
\end{eqnarray}
As discussed below, the limits on $s_{34}$ do not depend sensitively
on this choice.  The lepton masses, which have relatively little
effect on the limit on $s_{34}$, are chosen as $m_{\nu_4}=100$ GeV and
$m_{l_4}=145$ GeV.\footnote{We assume a Dirac mass for $\nu_4$. A
  dynamically generated Majorana mass for the fourth neutrino which
  made a negative contribution to $T$ could weaken the
  constraints for given \mtp.\cite{bhnu}}

Table 1 shows that the three large-mixing parameter sets of
\cite{blrr} are excluded ``with extreme prejudice'' by the EW
data. The large contributions to $T_4$ are responsible for the huge
\chisqsp values. In these fits the dominant contributors to
\chisqsp are the $W$ boson mass, the $Z$ boson width, and one or both
of \afbb and \alr. Since the FCNC constraints for these three 
parameter sets have been carefully considered in \cite{blrr}, 
we include them in the 95\% CL limit fits presented below. 
 
\begin{table}
\begin{center}
\vskip 12pt
\begin{tabular}{c||ccc|c|ccccc}
\hline
\hline
 & $ m_{t^{\prime}}$ &$|s_{34}|$&$T_4$&\chisqsp &$m_W$&$\Gamma_Z$&\afbb
       &\alr&$R_b$\\
\hline
I & 326 & 0.51& 1.09& 188&9&5&5&0.8&3\\
II&654&0.37& 3.58 &5750&53&30&27&24&7\\
III&389&0.63&2.59&2530&35&19&18&15&6\\
\hline
\hline
\end{tabular}
\end{center}
\caption{The three large-mixing parameter sets of \cite{blrr} with the
  corresponding values of $T_4$ from equation (3) and the \chisqsp for
  12 degrees of freedom from the global fits. The pulls of four
  sensitive observables are compared with the pull of $R_b$.}
\end{table}

Following the procedure of the EWWG,\cite{ewwg} the global fits are
obtained by minimizing \chisqsp while varying four SM parameters:
$m_t$, $\Delta \alpha_5$, $\alpha_S$, and \mh, with \mhsp allowed to
vary freely between 10 GeV and 1 TeV. Like the EWWG we leave
$\alpha_S$ unconstrained and determine it from the fits; for all the fits
at or within the 95\% CL limit for $|s_{34}|$, the values of $\alpha_S$
are in reasonable agreement with other determinations.  The global
constraints on \th34sp are fairly valued because they allow for the
possibility of \chisqsp minima at points in the parameter space that
are quite different than the location of the SM minimum, and they are
efficient because they aggregate the effect of the new physics on all
of the relevant observables.

Radiative corrections for the global fits are computed with
ZFITTER\cite{zfitter}, including the two loop corrections to
\sinthlsp\cite{xw2loop} and $m_W$.\cite{mw2loop} Our fits include the
largest experimental correlations, taken from the EWWG.  When we use
the same measurement set, our SM fit agrees closely with the
EWWG. Unlike typical fits with the oblique parameters $S$ and $T$,
which are performed with respect to fixed reference values of the SM
parameters, we use the complete set of radiative corrections from
ZFITTER to compute the dependence of \chisqsp on the SM parameters
$m_H$, $m_t$, $\Delta \alpha_5$, and $\alpha_S$, for {\em each}
($S_4,\, T_4$) pair, which represent only the fourth family
corrections. This procedure is then more accurate, however the resulting
$S,\, T$ values are not directly comparable to $S$ and $T$ from
typical fits with fixed ``reference'' values of the SM parameters.

In this work we focus on the EWWG set of observables.\footnote{ Unlike
  the EWWG we do not include the $W$ boson width, which with a 2\%
  error is not a precision measurement in the sense of the other
  measurements that typically have part per mil precision. In any
  case, $\Gamma_W$ has a negligible effect on the fits.} We also
briefly describe the results for the data set without the three
hadronic front-back asymmetry measurements.\cite{mcmh} The fits are
performed for four-family models with \mtpsp between 300 GeV and 1
TeV, including the three parameter sets of \cite{blrr}, with \bpsp and
lepton masses as described above. In all cases the \chisqsp minimum
occurs at $\theta_{34}=0$. The 95\% CL upper limit is obtained by
increasing \th34sp until \chisqsp increases by 3.84 units,
corresponding to CL$(\Delta \chi^2,1)=0.95$, like the procedure
commonly used to obtain the 95\% upper limit on \mhsp in the SM
fits.\footnote{For the SM fits the limit is at $\Delta \chi^2=2.71$
  corresponding to the 90\% symmetric confidence interval and the 95\%
  upper limit. Because the bound on $|\theta_{34}|$ is one-sided, the
  95\% limit is at $\Delta \chi^2=3.84$.} The fits at
$m_{t^{\prime}}=1\, {\rm TeV}$ are intended primarily to probe the range
of applicability of perturbation theory.

\begin{figure}
\centerline{\includegraphics[width=4in,angle=90]{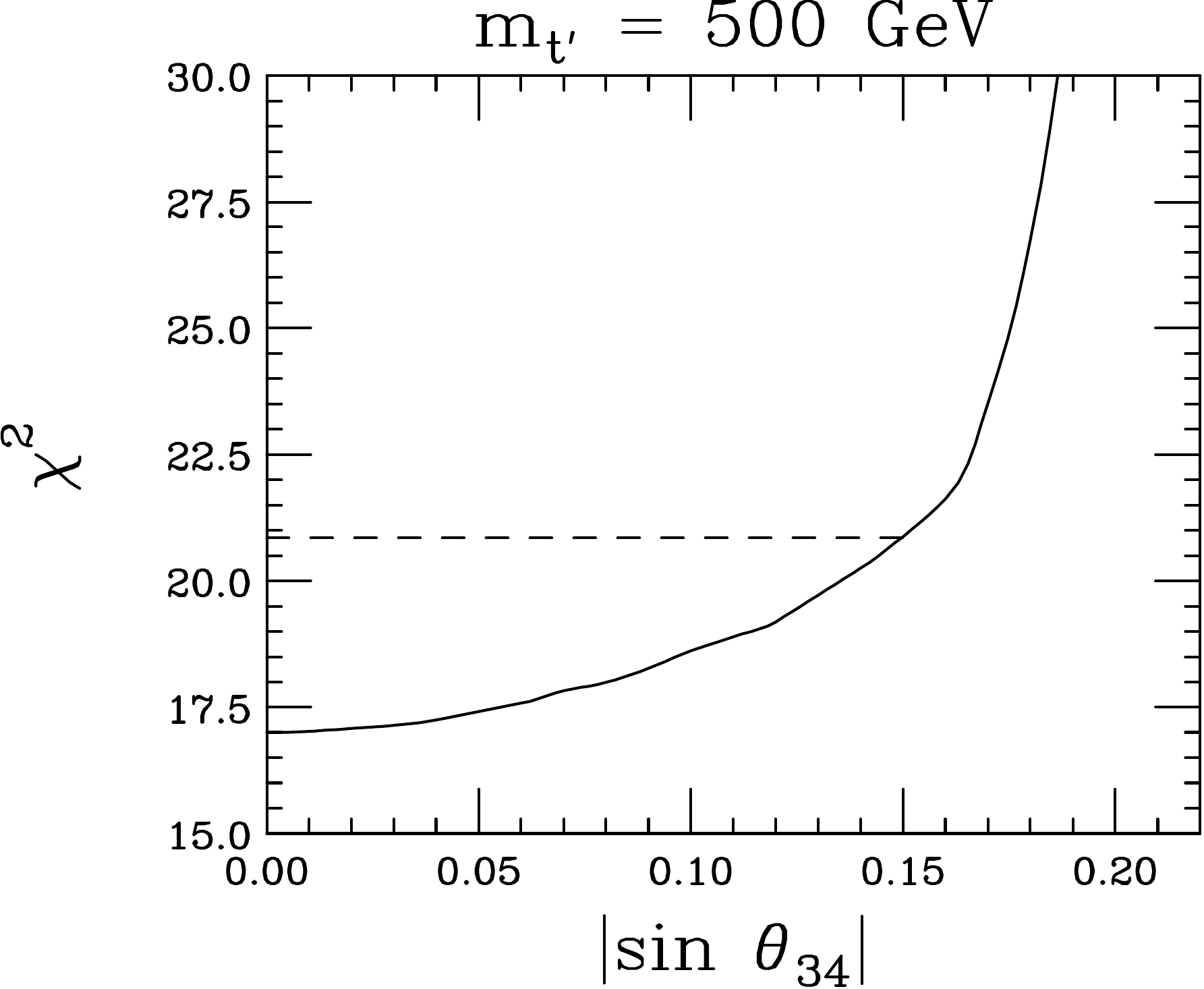}}
\caption{$\chi^2$ distribution as a function of $|{\rm
    sin}\theta_{34}|$ for the global fit to the four family model with
  $m_{t^{\prime}}= 500$ GeV. The horizontal line indicates the 95\% 
confidence interval. }
\label{fig1}
\end{figure}

\begin{table}
\begin{center}
\vskip 12pt
\begin{tabular}{c|c|cc|cc|cc}
\hline
\hline
 &Experiment& {\bf SM} & Pull &{\boldmath ${\rm SM}_4$} & Pull
      & {\boldmath $s_{34}[95\%]$} & Pull \\ 
\hline
$A_{LR}$ & 0.1513 (21)  & 0.1480  & 1.6& 0.1466  & 2.2&0.1457&2.7  \\
$A_{FB}^l$ & 0.01714 (95) &0.01642  & 0.8&0.1612  &1.1&0.01592&1.3  \\
$A_{e,\tau}$ & 0.1465 (32) & 0.1480 & -0.5& 0.1466 & -0.03&0.1457&-0.3 \\
$A_{FB}^b$ & 0.0992 (16) & 0.1037 & -2.8&0.1028&-2.2&0.1021&-1.8   \\
$A_{FB}^c$ & 0.0707 (35) & 0.0741 & -1.0&0.0734&-0.8&0.0729&-0.6  \\
$Q_{FB}$ & 0.23240 (120) & 0.23140 &-0.8&0.23158&-0.7&0.23169&-0.6  \\
$m_W$ & 80.398 (25) & 80.374 & 0.9 & 80.398 & 0.0&80.413&-0.6 \\
$\Gamma_Z$ & 2495.2 (23) & 2495.9 &0.3& 2498.1 & -1.3&2498.8&-1.6  \\
$R_{\ell}$ & 20.767 (25) &20.744  & 0.9 &20.733 & 1.4& 20.726&1.6  \\
$\sigma_h$ & 41.540 (37) & 41.477 &1.7  & 41.484 &1.5&41.487&1.4  \\
$R_b$ & 0.21629 (66) & 0.21586 &0.7& 0.21587 &0.6&0.21547&1.2  \\
$R_c$ & 0.1721 (30) & 0.1722 &-0.04& 0.1722 &-0.03&0.1723&-0.07  \\
$A_b$ & 0.923 (20) & 0.935 &-0.6 & 0.935 &-0.6&0.934&-0.6\\
$A_c$ & 0.670 (27) &  0.668 & 0.07&  0.668& 0.09&0.667&0.07 \\
$m_t$ & 172.6 (1.4) &172.3  &0.2&172.3  &0.2&172.3  &0.2  \\
$\Delta \alpha_5(m_Z)$ & 0.02758 (35) &0.02768& -0.3&0.02747& 0.3 
               &0.2732 & 0.7\\
$\alpha_S(m_Z)$ &    &0.1186& &0.1174&&0.1186&  \\
\hline
\mtp &&  & &500& &500& \\
$s_{34}$ &&&&0.0&& 0.15 & \\
$T_4$ &&&&0.20&&0.48&\\
$S_4$ &&&&0.15&&0.15&\\
$x_{t^{\prime}}$ &&&&0.0&&0.00052&\\
\hline
$m_H$ & & 85 && 139 &&810&\\
CL$(m_H > 114)$ & & 0.26 && 0.67 &&1.00& \\
$m_H(95\%)$&&148&&235& &$> 1000$&\\
\hline
$\chi^2$/dof& & 17.3/12 && 17.0/12 &&20.9/12& \\
CL($\chi^2)$ & & 0.14 &&0.15 &&0.05& \\
\hline
\hline
\end{tabular}
\end{center}
\caption{Global fits: the SM, the 4 family SM with $m_{t^{\prime}}=500$ 
GeV and $s_{34}=0$ and again with $s_{34}$ at the 95\% confidence level.}
\end{table}

The results for $m_{t^{\prime}}=500\, {\rm GeV}$ are illustrated in
figure 1 and table 2. Table 2 displays the SM fit, the four-family fit
for $m_{t^{\prime}}=500$ GeV and $\theta_{34}=0$ and also for \th34sp
at its (one loop) 95\% upper limit, $|s_{34}|=0.15$. \chisqsp as a function of
$|s_{34}|$ is shown in figure 1. The \chisqsp of the four family model
with $\theta_{34}=0$ is little changed from the SM but the central
value of \mhsp is increased, from 85 to 139 GeV, and the $3.2\sigma$
discrepancy between \alrsp and \afbbsp is more equally shared between
the two, in contrast to the SM fit in which \afbbsp is the
outlier.\footnote{ We assign 12 degrees of freedom to each four-family
  fit, assuming discovery at the LHC as a prior. This convention has
  no effect on the 95\% CL limit.} In the fit at the edge of the 95\%
confidence level, \mhsp increases to the strong coupling regime near 1
TeV because of the increased value of $T_4$, and \alrsp has become the
outlier, while the pull of $R_b$ is only 1.2. The central value of
\mhsp as a function of \sth34sp is shown in figure 2.

\begin{figure}
\centerline{\includegraphics[width=4in,angle=90]{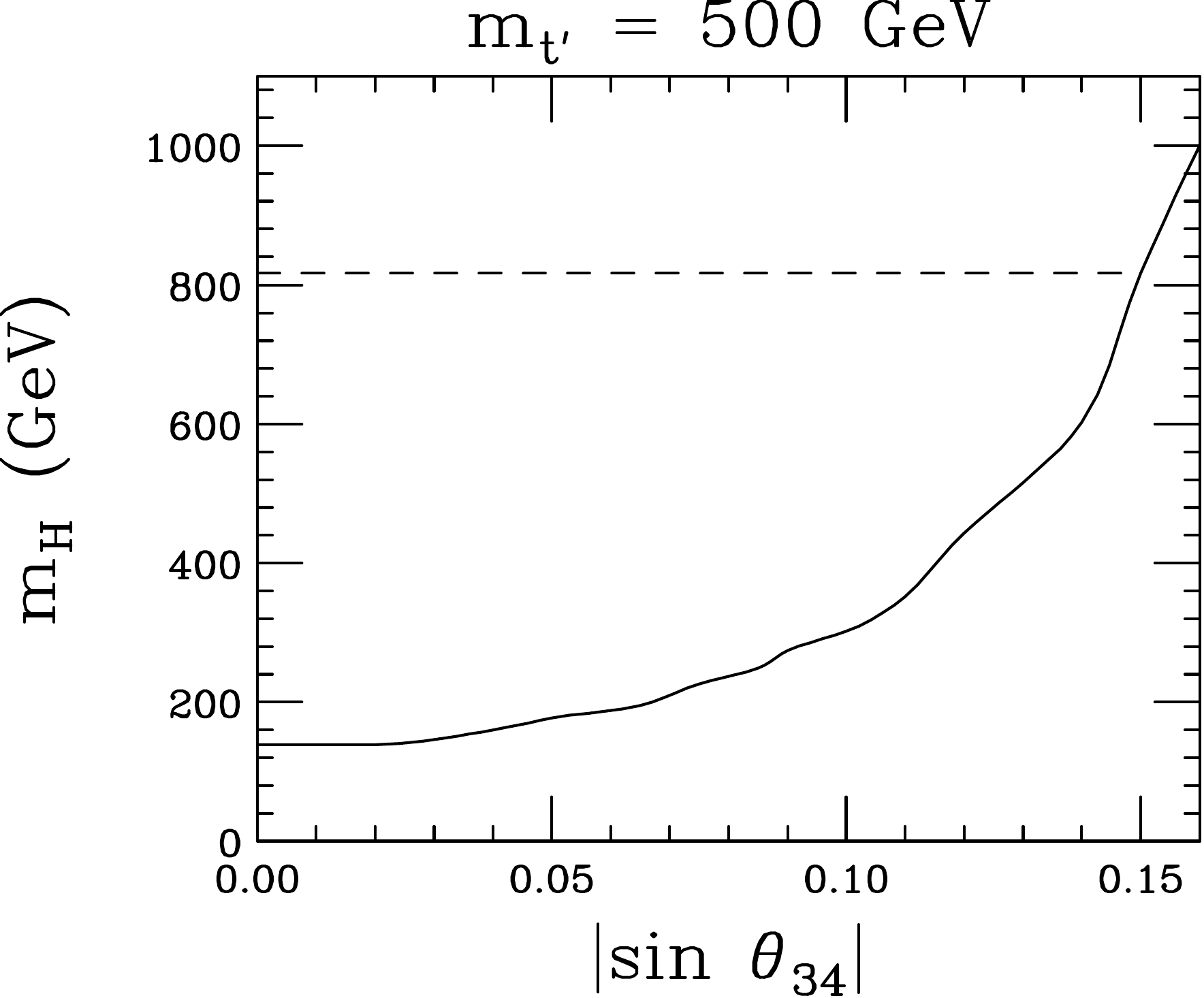}}
\caption{Higgs boson mass as a function of $|{\rm sin}\theta_{34}|$
  for the global fit to the four family model with $m_{t^{\prime}}=
  500$ GeV. The horizontal line indicates the 95\% confidence interval
  for $|{\rm sin}\theta_{34}|$. }
\label{fig2}
\end{figure}

In table 3 we summarize the limits on \th34sp for seven values of
\mtpsp between 300 to 650 GeV, including the three parameter sets from
reference \cite{blrr}, and in addition at $m_{t^{\prime}}=1\, {\rm
  TeV}$.  In all cases the fits at $\theta_{34}=0$ are nearly
identical to the fit shown in table 2 for $m_{t^{\prime}}= 500$ GeV.
As we would expect from equations (3) and (7), the limit on \sth34sp
becomes proportional to $1/m_{t^{\prime}}$ for $m_{t^{\prime}} \gg m_t$.
For these fits at the 95\% confidence limit, the Higgs boson
mass is at $m_H= 790 \pm 30$ GeV and $T_4 = 0.47 \pm 0.01$.  In all
cases we have $|V_{tb}|\simeq |V_{t^{\prime}b^{\prime}}|\simeq |{\rm
  cos}\theta_{34}| \geq 0.94$ and for $m_{t^{\prime}}\geq 500$ GeV we
have $|{\rm cos}\theta_{34}|\geq 0.99$.

\begin{table}
\begin{center}
\vskip 12pt
\begin{tabular}{c|cc|ccc}
\hline
\hline
  $ m_{t^{\prime}}$&$T_4$ &\mh(GeV)&$|s_{34}^{(1)}|$&
      $|s_{34}^{(2)}|\pm \Delta^{(2)}_{tb^{\prime}}$&$|c_{34}^{(2)}|$\\
\hline
300& 0.46&760&0.32&$0.35 \pm 0.001$& 0.94\\
326 & 0.47&760&0.28&$0.30 \pm 0.002$& 0.95\\
389 & 0.48&760&0.21 &$ 0.23 \pm 0.004$&0.97\\
400 & 0.47&800&0.20 & $0.22 \pm 0.005$&0.98\\
500 & 0.48&810&0.15 & $0.17 \pm 0.007$&0.99\\
600 & 0.48&800&0.12 & $0.14 \pm 0.010$&0.99\\
654 & 0.48&820&0.11 & $0.13 \pm 0.013$&0.99\\
1000 & 0.49&820&0.07 & $0.11 \pm 0.10$&0.99\\
\hline
\hline
\end{tabular}
\end{center}
\caption{95\% CL upper limits on $|s_{34}|$ at one and two loops
  from global fits to the EWWG data set. $T_4$ and \mhsp from the 95\%
  CL fits are also shown.}
\end{table}

To gauge the range of applicability of perturbation theory for large
quark masses, the limits in table 3 have been obtained at both one and
two loop order in the leading, nondecoupling electroweak corrections
to the $\rho$ parameter.\footnote{We have not included the two loop
  correction to the $Z\overline bb$ vertex, since the nondecoupling
  vertex correction does not have an important effect on the global
  fits.}  The leading two loop corrections for large quark mass are
known,\cite{vdb,rb,tar} but in no case with both the Higgs
boson mass dependence, which is large, {\it and} the dependence on the
masses of both quarks in the doublet.  The two loop corrections
computed by Barbieri {\it et al.},\cite{rb} are best suited for our
purpose: they include the full \mhsp dependence but for only one heavy
quark in the doublet, i.e., for $m_{Q1}\gg m_{Q2}\simeq 0$. This captures the
contribution that is most important for the bound on the mixing angle,
because the ${\rm sin}\theta_{34}$ dependent term in $T_4$, equation
(3), is dominated by $F_{t^{\prime}b}$, for which it is always an 
excellent approximation. 

We do not need to consider the two loop correction to
$F_{t^{\prime}b^{\prime}}$, because even the one loop term has a
negligible effect on the ${\rm sin}\theta_{34}$ dependence of $T_4$.
The $F_{tb^{\prime}}$ term is however somewhat problematic. The two
loop correction, $F^{(2)}_{tb^{\prime}}$, can safely be neglected for
the smallest masses we consider, say $m_{t^{\prime}}\, \ltap\, 400\,
{\rm GeV}$, for which the one loop term $F^{(1)}_{tb^{\prime}}$ is not
very important, but for $500\, {\rm GeV}\, \ltap\, m_{t^{\prime}}\,
\ltap\, 1\, {\rm TeV}$, $F^{(1)}_{tb^{\prime}}$ is not negligible and
approximating $m_{b^{\prime}}=0$ in the two loop correction
$F^{(2)}_{tb^{\prime}}$ may be a poor approximation. To give a
conservative indication of the possible error, the effect on the
bounds of a $\pm 100\%$ variation in the value of
$F^{(2)}_{tb^{\prime}}$ is shown as $\pm \Delta^{(2)}_{tb^{\prime}}$
in table 3. The conclusion is that the bounds on ${\rm
  sin}\theta_{34}$ are probably good to a few percent at
$m_{t^{\prime}}=300\, {\rm GeV}$, while at $m_{t^{\prime}}=650\, {\rm
  GeV}$ they are probably valid to from 10 to 20\%. At
$m_{t^{\prime}}=1\, {\rm TeV}$ the order one variation from
$\Delta^{(2)}_{tb^{\prime}}$ significantly overestimates the
uncertainty from $F^{(2)}_{tb^{\prime}}$, but the order 50\% shift
from the first to second order limit on $|s_{34}|$, from 0.07 to 0.11,
is a signal that perturbation theory has become unreliable.

\begin{figure}
\centerline{\includegraphics[width=4in,angle=90]{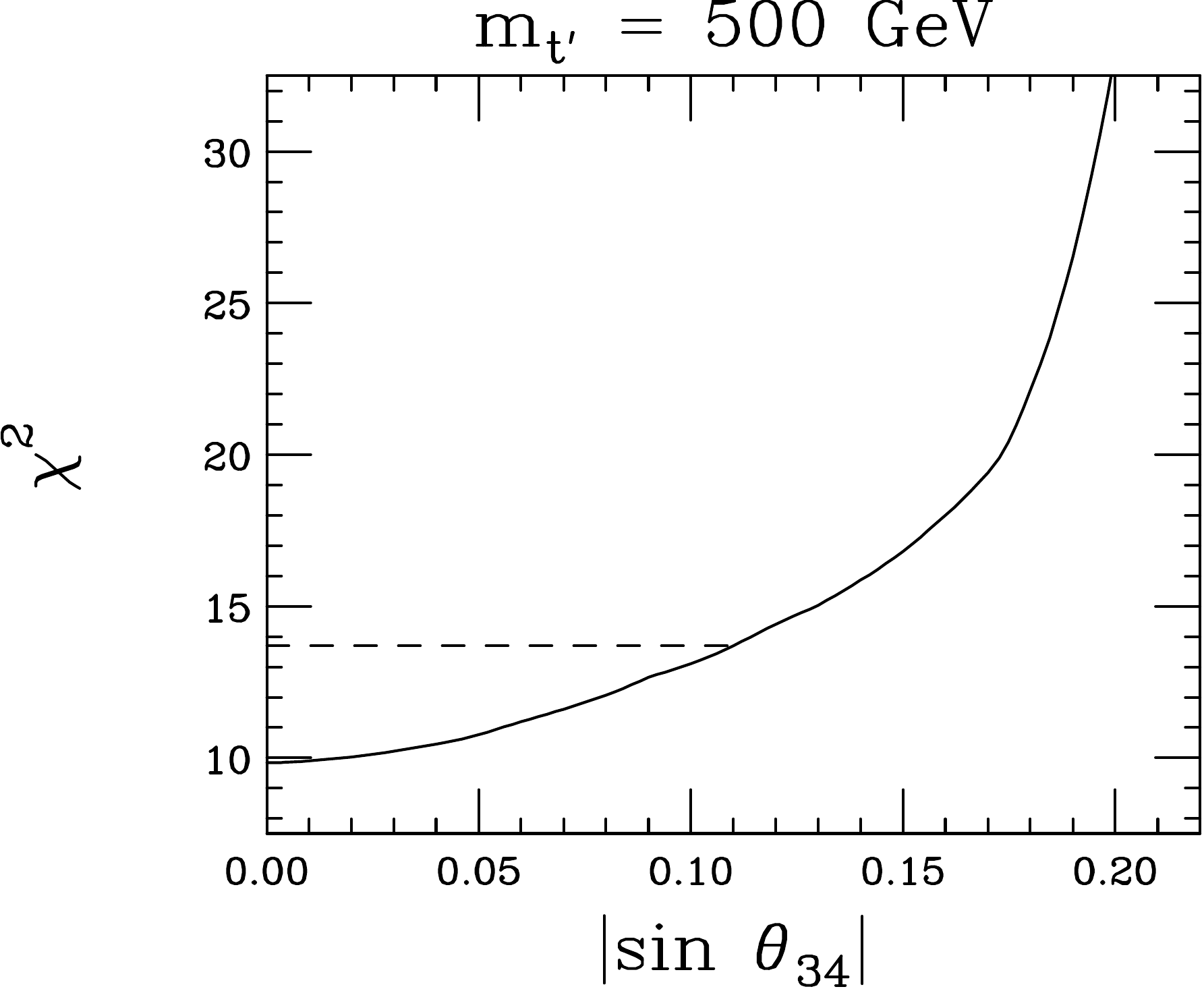}}
\caption{$\chi^2$ distribution as a function of $|{\rm
    sin}\theta_{34}|$ of the global fit to the four-family model with
  $m_{t^{\prime}}= 500$ GeV, for the data set without the hadronic
  asymmetry measuements. The horizontal line indicates the 95\%
  confidence interval. }
\label{fig3}
\end{figure}

The upper limits do not depend sensitively on the choice of
$m_{t^{\prime}}-m_{b^{\prime}}$ in equation (8). For larger mass
differences, e.g., $m_{t^{\prime}}-m_{b^{\prime}}\, \gtap\, 100\, {\rm
  GeV}$, there are no aceptable fits: at $\theta_{34}=0$ the fits are
poor, with ${\rm CL}(\chi^2)< 0.03$, and quickly become much poorer as
$|\theta_{34}|$ increases. For smaller mass splitting, e.g., at the
extreme, $m_{t^{\prime}}=m_{b^{\prime}}$, the \chisqsp CL's are
acceptable for $\theta_{34}=0$, with ${\rm CL}(\chi^2)= 0.18$, but the
predictions for the Higgs mass are unacceptable, with $m_H=35\, {\rm
  GeV}$ and ${\rm CL}(m_H > 114) = 0.0016$. In this case marginally
acceptable fits can be found by increasing $|s_{34}|$, which raises
the Higgs mass toward the allowed region while maintaining acceptable
\chisqsp CL's. For instance, with $m_{t^{\prime}}=m_{b^{\prime}}=500\,
         {\rm GeV}$ and $s_{34}=0.07$, the fit is just consistent at
         5\% CL with the 114 GeV lower limit on \mh, with $m_H=58\,
         {\rm GeV}$, ${\rm CL}(m_H > 114) = 0.05$ and ${\rm
           CL}(\chi^2)\simeq 0.18$. Taking $\Delta \chi^2=3.84$ from
         either this fit or from the fit at $\theta_{34}=0$, we find
         the one loop 95\% upper limit at $|s_{34}|<0.155$, little
         changed from the limit that we obtained using
         $m_{t^{\prime}}-m_{b^{\prime}} = 55\, {\rm GeV}$. There is a
         simple reason for the insensitivity of the limit to the mass
         splitting: as $|s_{34}|$ inceases, the terms in equation (3)
         that control the limit are $F_{t^{\prime}b}$ and
         $F_{tb^{\prime}}$ while $F_{t^{\prime}b^{\prime}}$, which is
         sensitive to $m_{t^{\prime}}-m_{b^{\prime}}$, plays a
         relatively minor role.

It is interesting to consider the data set with hadronic asymmetries
excluded, motivated by the possibility that underestimated sytematic
error might contribute to the $3.2\sigma$ discrepancy in the SM
determination of \sinthlsp from the hadronic and leptonic asymmetry
measurements.\cite{mcmh} The hadronic asymmetry measurements are more
challenging, both experimentally and, especially, theoretically,
because of the difficulty of extracting quark asymmetries from the
actual measurements of hadronic final states. The QCD corrections are
large, three times the experimental error, and the Monte Carlo
calculations needed to merge the large QCD corrections with the
experimental cuts give rise to a systematic error that is difficult to
quantify. Without the three hadronic measurements, \afbb, \afbc, and
\qfb, the confidence level of the SM fit improves dramatically but the
fit predicts a very light Higgs boson, $m_H=50$ GeV, with 95\% upper
limit at 105 GeV, in conflict at 97\% CL with the 114 GeV LEP II lower
limit. Results are shown in figure 3 and tables 4 and 5. As found by
Novikov {\it et al.}\cite{okun}, the four-family model removes the
conflict with the LEP II lower limit for this data set.  The limits in
table 4 are $\simeq 40\%$ stronger than for the full data set in table
3. At the 95\% limit for \th34sp the Higgs boson masses range from 275
to 300 GeV with $T_4=0.35$ in all cases.  Because $T_4$ and \mhsp are
smaller for this data set, both the one loop and two loop corrections
are smaller, and perturbation theory appears to be under better
control.

\begin{table}
\begin{center}
\vskip 12pt
\begin{tabular}{c|cc|ccc}
\hline
\hline
  $ m_{t^{\prime}}$&$T_4$ &\mh(GeV)&$|s_{34}^{(1)}|$&
      $|s_{34}^{(2)}|\pm \Delta^{(2)}_{tb^{\prime}}$&$|c_{34}^{(2)}|$\\
\hline
300& 0.35&300&0.25&$0.26 \pm 0.0008$& 0.97\\
326 & 0.35&280&0.21&$0.22 \pm 0.0010$& 0.98\\
389 & 0.35&270&0.16 &$ 0.17 \pm 0.0016$&0.99\\
400 & 0.35&290&0.15 & $0.16 \pm 0.0016$&0.99\\
500 & 0.35&270&0.11 & $0.12 \pm 0.0027$&0.99\\
600 & 0.35&290&0.087 & $0.095 \pm 0.0033$&0.995\\
654 & 0.35&280&0.078 & $0.086 \pm 0.0035$&0.996\\
1000 & 0.35&270&0.048 & $0.059 \pm 0.007$&0.998\\
\hline
\hline
\end{tabular}
\end{center}
\caption{95\% CL upper limits on $|s_{34}|$ at one and two loops
  from global fits to the EWWG data set without the hadronic asymmetry
  measurements. $T_4$ and \mhsp from the 95\% CL fits are also shown.}
\end{table}

\begin{table}
\begin{center}
\vskip 12pt
\begin{tabular}{c|c|cc|cc|cc}
\hline
\hline
 &Experiment& {\bf SM} & Pull &{\boldmath ${\rm SM}_4$} & Pull
      & {\boldmath $s_{34}[95\%]$} & Pull \\ 
\hline
$A_{LR}$ & 0.1513 (21)  & 0.1503 & 0.5& 0.1483 & 1.4&0.1474&1.8  \\
$A_{FB}^l$ & 0.01714 (95) &0.01694 & 0.2&0.1649 &0.7&0.01630&0.9  \\
$A_{e,\tau}$ & 0.1465 (32) & 0.1503 & -1.2& 0.1483 & -0.6&0.1474&-0.3 \\
$m_W$ & 80.398 (25) & 80.403& 0.03 & 80.423 & -1.0&80.425&-1.1 \\
$\Gamma_Z$ & 2495.2 (23) & 2496.0 &-0.3& 2498.5 & -1.4&2499.2&-1.7  \\
$R_{\ell}$ & 20.767 (25) &20.741 & 1.0&20.729 & 1.5& 20.725&1.7  \\
$\sigma_h$ & 41.540 (37) & 41.482&1.6 & 41.489 &1.4&41.491&1.3  \\
$R_b$ & 0.21629 (66) & 0.21584 &0.7& 0.21586 &0.6&0.2157&1.0  \\
$R_c$ & 0.1721 (30) & 0.1722 &-0.04& 0.1722 &-0.03&0.1722&-0.05  \\
$A_b$ & 0.923 (20) & 0.935 &-0.6 & 0.935 &-0.6&0.935&-0.6\\
$A_c$ & 0.670 (27) &  0.669 & 0.03&  0.668& 0.06&0.668&0.08 \\
$m_t$ & 172.6 (1.4) &172.3  &0.2&172.3  &0.2&172.3  &0.2  \\
$\Delta \alpha_5(m_Z)$ & 0.02758 (35) &0.02754& 0.1&0.02747& 0.3 
               &0.2732 & 0.7\\
$\alpha_S(m_Z)$ &    &0.1174& &0.1162&&0.1168&  \\
\hline
\mtp &&  & &500& &500& \\
$s_{34}$ &&&&0.0&& 0.11 & \\
$T_4$ &&&&0.20&&0.35&\\
$S_4$ &&&&0.15&&0.15&\\
$x_{t^{\prime}}$ &&&&0.0&&0.00028&\\
\hline
$m_H$ & & 50 && 89 &&280&\\
CL$(m_H > 114)$ & & 0.03 && 0.28 &&1.0& \\
$m_H(95\%)$&&105&&174& &$480$&\\
\hline
$\chi^2$/dof& & 5.6/9 && 9.8/9 &&13.7/9& \\
CL($\chi^2)$ & & 0.78 &&0.36 &&0.13& \\
\hline
\hline
\end{tabular}
\end{center}
\caption{Global fits for the data set without the hadronic asymmetry
  measurements: the SM, the 4 family SM with $m_{t^{\prime}}=500$ GeV
  and $s_{34}=0$, and again with $s_{34}$ at the 95\% confidence
  level.}
\end{table}

{\it \noindent \underline{Discussion}}

While this work was initially motivated by the paper of Bobrowski {\it
  et al.}, several other studies have considered the possible role of
a fourth generation on FCNC phenomena and the CKM matrix.  The fits of
Yanir\cite{yanir} for $m_{t^{\prime}}=500\, {\rm GeV}$ require
$|s_{34}|\, \ltap\, 0.14$ and therefore fall within the 95\% CL limit
of the fit to the complete data set, table 3, and just beyond the
limit from the fit with hadronic asymmetries excluded, table 4.  The
95\% CL limit quoted by Alwall {\it et al.},
$|c_{34}|>0.93$,\cite{alwall} is weaker than the global fit limits
in tables 3 and 4, especially for the larger values of \mtp.
Herrera {\it et al.}\cite{herrera} have studied the FCNC constraints
in texture models of the $4\times 4$ quark mass matrices. They remark
on an isolated solution with $|V_{tb}| \sim 0.88$, implying
$|s_{34}|\sim 0.47$, which is decisively excluded by the EW fits, like
the parameter sets in table 1. Most of their fits lie within $0.90
\leq |V_{tb}| \leq 0.94$, which is excluded at 95\% CL in all cases
considered here except the fit to the complete data set with
$m_{t^{\prime}}=300\, {\rm GeV}$ shown in table 3.

Hou {\it et al.}\cite{hou} and Soni {\it et al.}\cite{soni} have
identified regions in the CKM$_4$ parameter space that could explain
possible anomalies in B meson CP measurements, requiring large mixing
between not only the third and fourth families but also between the
second and the fourth. For instance, Hou {\it et al.} consider the
four family model with $m_{t^{\prime}}=300\, {\rm GeV}$, $s_{34}=
0.22$, and with 2-4 mixing given by $|V_{t^{\prime}s}|=0.114$ and
$|V_{cb^{\prime}}|=0.116$. There are then nonnegligible contributions
to $T_4$ from the 2-4 family mixing and additional terms must be added
to equation (3), which becomes 
\begin{eqnarray}
T_4 &=& \frac{1}{8\pi x_W(1-x_W)}\left\{ 3\left[
     F_{t^{\prime}b^{\prime}} +s_{34}^2(F_{t^{\prime}b} 
     +F_{tb^{\prime}} -F_{tb} -F_{t^{\prime}b^{\prime}})\right]
     + F_{l_4\nu_4}\right\}  \nonumber \\
   & & +\ \frac{3}{8\pi x_W(1-x_W)}\left[|V_{t^{\prime}s}|^2 F_{t^{\prime}s} 
     + |V_{cb^{\prime}}|^2 F_{cb^{\prime}}\right].
\end{eqnarray}
Using the above mixing angles and $m_{t^{\prime}}=300\, {\rm GeV}$ the
fit to the full data set yields $\Delta \chi^2 = 2.44$, which falls 
within the 95\% CL limit at $\Delta \chi^2 = 3.84$. For the fit 
with hadronic asymmetries excluded we find $\Delta \chi^2 = 4.64$, just 
beyond the 95\% CL limit. 

Soni {\it et al.}\cite{soni} quote a range of values for the product
$\lambda^s_{t^{\prime}} = |V^*_{t^{\prime}s} V_{t^{\prime}b}|$, for
values of $m_{t^{\prime}}$ between 400 and 700 GeV. The limit on
$\lambda^s_{t^{\prime}}$ then depends on the hierarchy between the CKM
matrix elements,
\begin{eqnarray}
 r=\frac{|V_{t^{\prime}s}|}{|V_{t^{\prime}b}|}.
\end{eqnarray}
Neglecting a possible contribution which might be expected from
$F_{cb^{\prime}}$ but is not specified in \cite {soni}, the limits
quoted in tables 3 and 4 for $s_{34}^2$ now apply instead to the combination
$(1+r^2)s_{34}^2$.  The corresponding upper limit on
$\lambda^s_{t^{\prime}} = rs_{34}^2$ is then a function of $r$,
\begin{eqnarray}
 \lambda^s_{t^{\prime}} < \frac{r}{1+r^2}\, X_{95}
\end{eqnarray}
where $X_{95}$ is the square of the 95\% CL upper limits on $s_{34}$
given in the tables. The limit is maximal for $r=1$, corresponding to
$\lambda^s_{t^{\prime}}< X_{95}/2$; in this case the full range of
preferred values in \cite{soni} is allowed for the fit to the full
data set and only slightly restricted in the fit with hadronic
asymmetries excluded. If we assume a hierarchy of order the Cabibbo
angle, $r \sim 0.2$, then the bound is tighter,
$\lambda^s_{t^{\prime}}\, \ltap\, X_{95}/5$; in this case a portion of
the preferred range in \cite{soni} is excluded at 95\% CL for both
data sets, but a significant portion continues to be allowed.

If a fourth family were discovered at the LHC, the subsequent study of
its properties would be a major undertaking, with many profound
implications. The elucidation of the four family CKM matrix would be
important to understand the on-shell measurements at the LHC as well
as the virtual implications for flavor physics and CP
violation. Electroweak precision data, perhaps eventually augmented by
a second generation $Z$ boson factory, should continue to play an
important role, by constraining and vetting the emerging picture and
even by indicating a mismatch with direct high energy measurements
that could be a signal for still unobserved new physics.

\vskip .2in
\noindent {\it Acknowledgements:} I would like to thank Zoltan Ligeti for 
several helpful discussions. 

\noindent{\small This work was supported in part by the Director, Office of
Science, Office of High Energy and Nuclear Physics, Division of High
Energy Physics, of the U.S. Department of Energy under Contract
DE-AC02-05CH11231}

\end{document}